\documentclass[12pt,preprint]{aastex}

\slugcomment{Submitted to ApJ  10/20/2011,  Accepted  12/16/2011}

\shorttitle{Star Formation Rates and Extragalactic Scaling Relations}
\shortauthors{Lada, Forbrich, Lombardi \& Alves}

\def\lsun{L$_\odot$}

\def\msun{M$_\odot$}

\def\cc{cm$^{-3}$}

\def\13co{$^{13}$CO}

\begin{document}


\title{Star Formation Rates in Molecular Clouds and the Nature of the Extragalactic Scaling Relations}

\author{Charles J. Lada }
\affil{Harvard-Smithsonian Center for Astrophysics, 60 Garden Street
Cambridge, MA 02138, USA}
\email{clada@cfa.harvard.edu}

\author{Jan Forbrich }
\affil{Harvard-Smithsonian Center for Astrophysics, 60 Garden Street
Cambridge, MA 02138, USA}
\email{jforbrich@cfa.harvard.edu}

\author{Marco Lombardi}
\affil{University of Milan, Department of Physics, via Celoria 16, 220133 Milan, Italy}
\email{marco.lombardi@gmail.com}

\and

\author{Jo\~ao F. Alves}
\affil{Institute for Astronomy,  University of Vienna, T\"urkenschanzstrasse 17, 1180 Vienna, Austria}
\email{joao.alves@univie.ac.at}

\begin{abstract}

In this paper we investigate scaling relations between star formation rates and molecular gas masses for both local Galactic clouds and a sample of external galaxies. We  specifically consider relations between the star formation rates and measurements of dense, as well as total, molecular gas masses.  We argue that there is a fundamental empirical scaling relation that directly connects the local star formation process with that operating globally within galaxies. Specifically, the total star formation rate in a molecular cloud or galaxy is linearly proportional to the mass of dense gas within the cloud or galaxy. This simple  relation, first documented in previous studies,  holds over a span of mass covering nearly nine orders of magnitude and indicates that the rate of star formation is directly controlled by the amount of dense molecular gas that can be assembled within a star formation complex. We further show that the star formation rates and total molecular masses, characterizing both local clouds and galaxies, are correlated over similarly large scales of mass and can be described by a family of linear star formation scaling laws, parameterized by $f_{DG}$, the fraction of dense gas contained within the clouds or galaxies. That is, the underlying star formation scaling law is always linear for clouds and galaxies with the same dense gas fraction. These considerations provide a single unified framework for understanding the relation between the standard (non-linear) extragalactic Schmidt-Kennicutt scaling law, that is typically derived from CO observations of the gas, and the linear star formation scaling law derived  from HCN observations of the dense gas.

\end{abstract}



\section{Introduction}

Knowledge of the physical factors that control the conversion of interstellar gas into stars is of fundamental importance for both developing a predictive physical theory of star formation and understanding the evolution of galaxies from the earliest epochs of cosmic history to the present time. An essential first step to obtaining such knowledge is to establish empirically the underlying relation or relationships that most directly connect the rate of star formation in a galaxy to some general physical property of the interstellar gas from which stars form. A little more than a half-century ago, Schmidt (1959) conjectured that this might take the form of a scaling relation between the rate of star formation and some power, n,  of the surface density of atomic (HI) gas. From evaluation of  the distributions of local HI gas and stars orthogonal to the Galactic plane, he suggested that n$\approx$ 2.  Subsequent studies comparing the surface densities of OB stars and HII regions with those of atomic gas within nearby external galaxies produced scaling laws with similar, super-linear, power-law indices (e.g., Sanduleak 1969; Hamajima \& Tosa 1975).

 By the 1980s it became clear that molecular, not atomic, clouds were the sites of star formation in galaxies.  The ability to make sensitive CO molecular-line observations enabled,  for the first time, the measurement of total  gas surface densities ($\Sigma_{HI + H_2}$) in external galaxies while advancements in infrared and ultraviolet observations  led to significant improvements in the measurements of star formation rates. Significant effort was then expended by a number of researchers to systematically measure star formation rates and total gas surface densities in increasingly large samples of galaxies (e.g., Kennicutt 1989 and references therein). These efforts culminated in the study of Kennicutt (1998a) who compiled galaxy averaged measurements of star formation rates and gas surface densities for a large sample of galaxies including normal spirals and starbursts. He derived a scaling relation between the star formation rate surface density ($\Sigma_{SFR}$) and total gas surface density ($\Sigma_{HI + H_2}$) that was characterized by a power-law index of n $\approx$ 1.4. This value was shallower than that Schmidt and others found for individual galaxies using only atomic gas but still super-linear.  Wong and Blitz ( 2002), employing spatially resolved observations of seven nearby, molecular rich, spiral galaxies, showed that the star formation rate was better correlated with the molecular hydrogen surface density, $\Sigma_{H_2}$, than with the atomic surface density, but still obtained n $\approx$ 1.4. More recently, Bigiel et al. (2008) analyzed spatially resolved observations of 18 nearby galaxies containing both atomic rich and molecular rich objects and confirmed that $\Sigma_{SFR}$ was better correlated with $\Sigma_{H_2}$ than $\Sigma_{HI}$, but they determined that n $=$ 1.0 ($\pm$ 0.2)  for the $\Sigma_{SFR}$ -- $\Sigma_{H_2}$ relation.  However, recent observations of M 101 and  M 81 have suggested that the index of the scaling law can vary within a galaxy with values of n ranging between 1 and 2 (Suzuki et al. 2010). 

Among the more interesting investigations of the extragalactic scaling laws for star formation was that of Gao and Solomon (2004) who used molecular-line emission from HCN, rather than CO, to trace the molecular gas. They found a linear (n $=$ 1) correlation between the total far-infrared luminosities and the HCN molecular-line luminosities of a large sample of star forming galaxies including normal spirals and luminous and ultra-luminous infrared galaxies.  Since the total infrared luminosity is a good proxy for the total star formation rate (SFR) and the HCN luminosity a good proxy for the total amount of dense (i.e., n(H$_2$) $\geq$ 3 $\times$ 10$^4$ cm$^{-3}$) gas in a galaxy, this also implied a linear correlation between the SFR and the mass of dense molecular gas. 
 
The various determinations of differing power-law indices for the extragalactic star formation scaling relations present a somewhat confused and problematic picture. Particularly since the difference between a linear and non-linear scaling relation can have significant consequences for the theoretical understanding of the star formation process in galaxies.  Therefore it is important to understand the nature of such differences. Are the different scaling relations consistent with each other? Are the differences due to such effects as the choice of the samples studied (e.g., normal spirals vs starbursts, CO rich vs. HI rich galaxies, distant vs. nearby systems, etc.) or the different quantities actually measured (e.g., SFR vs.  $\Sigma_{SFR}$, $\Sigma_{HI + H_2}$ vs. $\Sigma_{H_2}$, or CO vs. HCN, etc.), or the systematic uncertainties in the quantities measured (e.g., observational tracers or IMFs adopted for SFR determinations, conversion factor for transforming CO measurements into H$_2$ masses, etc.), or some linear combination of all these effects? Do any of these scaling relations represent the fundamental underlying physical relationship that most directly connects star formation activity with interstellar gas?

Schmidt's original scaling law was determined from observations of the local region of the Galaxy. Since our knowledge of the local Milky Way has improved profoundly over the last half century, it would seem that important insights into the relation between star formation and interstellar gas could and should be derived from observations of local star formation activity.
In a previous paper (Lada et al. 2010; hereafter Paper I) we presented a study of the star formation activity in a sample of local (d $<$ 0.5 kpc) molecular clouds with total masses between 10$^3$ and 10$^5$ \msun. We employed infrared extinction measurements derived from wide-field surveys to determine accurate cloud masses and mass surface densities, and compiled from the literature both ground and space-based infrared surveys of young stellar objects to construct complete inventories of star formation within the clouds of our local sample. We found the specific star formation rates (i.e., the star formation rates per unit cloud mass) in these clouds to vary by an order of magnitude, independent of total cloud mass. However, we also found the dispersion in the specific star formation rate,  to be minimized (and reduced by a factor of 2-3) if one considers only the mass of molecular gas characterized by high extinction in calculating the specific star formation rates. As a result we showed that the (total) star formation rate  in local clouds is linearly proportional to the cloud mass contained above an extinction threshold of A$_K$ $\geq$ 0.8 magnitudes, corresponding to a gas surface density threshold of $\Sigma_{H_2}$ $\approx$ 116 \msun pc$^{-2}$. Similar surface density thresholds for star formation in local clouds have been suggested in other recent studies (e.g., Goldsmith et al 2008; Heiderman et al. 2010). Given the density stratification of molecular clouds, we argued that such surface density thresholds also correspond to volume density thresholds of n(H$_2$) $\approx$ 10$^4$ cm$^{-3}$. These findings are consistent with and reinforce those of Wu et al. (2005) who had already demonstrated a linear correlation between far-infrared luminosity and HCN luminosity (i.e., between SFR and dense gas mass) for more massive and distant star formation regions in the Milky Way. 

The correspondence between these results and those obtained by Gao and Solomon (2004) for external galaxies is intriguing and especially striking because the scalings of the Galactic and extragalactic power-law relations, that together span more than nine orders of magnitude in cloud mass, agree to within a factor of 2-3. This suggested to us that the close relationship between the star formation rates and the {\it dense} gas masses of molecular clouds could be the underlying physical relation that connects star formation activity with interstellar gas over vast spatial scales from the immediate vicinity of the sun to the most distant galaxies.  

However, if this is so, how does one understand these observations in the context of the classical Schmidt-Kennicutt scaling relations based on CO observations?  These classical relations are often super linear and moreover, as Heiderman et al. (2010) point out, they  under predict the $\Sigma_{SFR}$ in local regions by factors of 17 - 50 (see also Evans et al. 2009). In this paper we  attempt to address this issue by re-examining the extinction observations of local clouds to include low extinction material and re-examining the CO observations of the clouds studied by Gao and Solomon. We show that all the observations can be understood within a self-consistent framework in which the differences are primarily due to the dense gas fractions that characterize the molecular gas being observed, supporting a hypothesis originally put forward  by Gao and Solomon (2004).

\section{The SFR-Molecular Mass Diagram}

\subsection{The Local Clouds}

In Figure 1 we plot the relation between the (total) star formation rate, SFR, and gas mass for the 11 clouds in the Paper I sample. The SFRs are from Table 2 of Paper I and are the averaged rates over a timescale of 2 Myrs. However, here we plot for each cloud two different masses derived from the infrared extinction measurements. The filled circles represent cloud masses measured above an infrared (K-band) extinction threshold of 0.8 magnitudes and correspond to the dense gas masses ($M_{DG}$) of the clouds. The open circles represent cloud masses measured above a lower infrared extinction threshold of  0.1 magnitudes and correspond to the total gaseous masses ($M_{TG}$) of the clouds. These latter masses should also approximately correspond to those that would be traced by CO emission, while the former masses approximately correspond to those that would be traced by HCN emission. The parallel dashed lines represent a series of linear relations between SFR and mass. The top line is the best fit linear relation for the high extinction (dense gas) masses (following Paper I). The two lower lines are the same relation only shifted or scaled in the horizontal direction by one and two orders of magnitude in mass, respectively. We can now express the star formation scaling law for these clouds as:

\begin{equation}
SFR \equiv \dot{M}_* = 4.6 \times 10^{-8} f_{DG} M_G (M_\odot)  \ \ \ M_\odot\ yr^{-1} \  \  \  
\end{equation}



\noindent
where  $M_{G}$ is molecular mass measured at a particular extinction threshold and corrected for the presence of Helium and $f_{DG}$ is the fraction of dense gas, i.e., $M_{DG} = f_{DG} M_G$. The three parallel lines correspond to $f_{DG}$ $=$ 1.0, 0.1 and 0.01 from left to right, respectively. These scalings essentially represent the fraction of the measured mass that is above the 0.8 magnitude extinction threshold or equivalently above a volume density threshold of roughly n(H$_2$) = 10$^4$ \cc (Paper I). These lines also correspond to lines of constant gas depletion times of  20 Myr, 200 Myr and 2 Gyr, respectively. For the open symbols on the plot, $M_G$ = $M_{TG}$, the total mass of the molecular cloud.  

The interesting aspect of this plot is that the low extinction (total) masses also appear to follow a linear scaling law, similar to that of the high extinction (high density) masses. Indeed, a formal least-squares fit to the former data produces a slightly sub-linear index value of 0.81 $\pm$ 0.19. The total cloud masses, $M_{TG}$, appear to follow and scatter around the relation given by Equation 1, if $f_{DG}$ $=$ 0.1. However the magnitude of the scatter around this linear relation is significantly higher than that for the high extinction masses around the best-fit line given by Equation 1 (i.e., $f_{DG}$ $=$ 1 and $M_G = M_{DG}$). Star formation occurs almost exclusively in gas characterized by high densities (n($H_2$) $>$ 10$^4$ \cc; Lada 1992) and the origin of the large scatter in the star formation scaling law for the total cloud masses is a direct result of the large variations in the dense gas (high extinction) fractions that are observed for these clouds (Paper I). In contrast to classical Schmidt-Kennicutt extragalactic scaling laws, there is no evidence for a super-linear scaling for the star formation law for local clouds, even when the total masses of the clouds are considered.

\subsection{Galaxies}
 
In order to compare galaxies with the galactic clouds on the SFR-Molecular Mass diagram we use the  sample of galaxies observed  by Gao and Solomon (2004; hereafter GS04).  Their sample consists of normal spirals and starburst galaxies, including luminous and ultra-luminous infrared galaxies (i.e., LIRGs \& ULIRGs). We selected this sample for comparison with our local cloud sample because it is the only sample of galaxies with systematically measured molecular masses using both a tracer of high density gas, HCN, and a tracer of total cloud mass, CO. In addition, the SFRs of the galaxies in the sample are all derived in the same manner from a homogeneous set of infrared observations. 

 It is a priori unclear whether the star formation rates and/or gas masses reported  for the GS04 galaxy sample are directly comparable to those reported in Paper I for the local cloud sample. The SFR for the local clouds was determined by direct counting of nearly complete inventories of Young Stellar Objects in each cloud and assuming a star formation timescale of 2 Myrs, while the SFRs for the GS04 galaxies are galaxy-wide averages that were derived from conversion of a FIR flux into a mass growth rate using stellar population synthesis models and assuming, among other parameters,  a simple Salpeter IMF and a timescale of 10-100 Myrs (Kennicutt 1998b).  Gao and Solomon use the most simple form of the virial theorem to convert HCN luminosity to a galaxy averaged dense gas mass, while in Paper I masses are calculated from direct integration of resolved extinction measurements of individual clouds and the assumption of a standard gas-to-dust ratio. Moreover, even if both mass calculations are accurate, it is not obvious that the A$_K$ $=$ 0.8 mag contour encompasses exactly the same mass as would be detected in HCN emission averaged over an entire cloud or galaxy.  We therefore would not necessarily expect the Galactic clouds and the Gao-Solomon galaxies to fall onto the exact same line  in the SFR-Molecular Mass diagram  (for dense gas masses) and, they do not.  Although previous studies (GS04, Paper I) independently found the relation between SFR and dense gas mass  to be linear for both local clouds and galaxies, the respective  coefficients (intercepts) differed by a factor of 2.7, with the galactic relation predicting higher SFRs for a given amount of dense gas. However, given the fact that these two linear relations together span nine orders of magnitude in mass, and their coefficients are consistent within the quoted errors (Paper I), it seems reasonable to conclude that they represent one and the same relation. 

Indeed, in a study of massive, but relatively distant, Galactic molecular clouds, Wu et al (2005) demonstrated a linear correlation between FIR and L$_{\rm HCN}$ for those clouds that was nearly identical (with similar coefficients) to that found by Gao and Solomon (2004). This finding thus extended the correlation between these two quantities over a range of more than 7-8 orders of magnitude and indicated that both the GS04 galaxies and Galactic clouds should lie on the same SFR-Mass relation for the dense gas component traced by HCN observations. Because the SFRs and masses calculated for the local sample are likely more robust than those determined for the Wu et al. clouds and the GS04 galaxies, we decided to adjust the coefficient of the GS04 relation to match that of the local sample for the dense molecular gas.  In principle, this could be accomplished by either, a) adjusting  the star formation rates upward,  b) adjusting the HCN masses downward, or c) simultaneously adjusting some appropriate linear combination of both these quantities. It is not obvious which of these alternatives would be most appropriate, and given the complexities and uncertainties in calculating both the star formation rates and dense gas masses for these galaxies, choice  c) might be the best option. However, for simplicity  we elected to match the coefficients by adjusting the GS04 star formation rates upward (by log($\Delta$SFR) = 0.43) so that they match those of Paper I when linearly extrapolated down to local cloud masses. {We emphasize here that the primary results and conclusions of this paper (see \S 3) are independent of the details (i.e., a, b or c) of how we choose to adjust the coefficient of the GS04 relation to match the relation for local clouds.} 
 
 In Figure 2 we extend the SFR-Molecular Mass plot to scales that can include measurements of  entire galaxies and we plot the galaxies in the GS04 sample. As with the local sample we plot two sets of masses for the GS04 galaxies. Again, the filled symbols correspond to dense gas masses, as measured using HCN emission. The open symbols correspond to total cloud masses measured from CO emission. 
The dense gas masses of the galaxies are those determined by GS04. Since GS04 did not report total gas (CO) masses for the galaxies in their sample, we made use of the CO(1-0) luminosities reported by GS04 to derive the total gas masses. We applied a conversion of $M_{\rm gas}$/$L_{\rm CO}$ = 1.36 $\times$ $\alpha_{\rm G}$ with a Galactic value of $\alpha_{\rm G}=3.2$  $M_\odot$ (K km s$^{-1}$ pc$^2$)$^{-1}$ (e.g., Genzel et al. 2010). 
The star formation rates for these galaxies are those determined by GS04 after the upward adjustment described described above.


Adjusting the GS04 SFRs upward implicitly assumes that the SFRs determined from L$_{\rm FIR}$ underestimate the true star formation rates, at least when extrapolated to local clouds. In an attempt to assess this possibility we investigated the relation between L$_{\rm FIR}$ and SFR in the local cloud sample. In the local cloud sample of Paper I, the SFR is dominated by the Orion A and B molecular clouds which account for 67\% of the total SFR for all the clouds in the sample. Following the prescription of GS04 we used IRAS observations to determine the FIR luminosity of  a 100 pc diameter region encompassing both the Orion A and B clouds. We calculated the FIR luminosity  to be 5.4 $\times$ 10$^5$  \lsun.  Using the relation 
 $\dot M_{\rm SFR} \approx 2 \times 10^{-10} (L_{\rm IR}/L_\odot) \ M_\odot yr^{-1}$, (following GS04 and Kennicutt 1998b), this corresponds to SFR $=$ 1.1 $\times$ 10$^{-4}$ \msun yr$^{-1}$, a value which is a factor of 8 lower than the combined SFR (8.7 $\times$ 10$^{-4}$ \msun yr$^{-1}$) determined for the Orion A and B clouds in Paper I.  We note that much of this deficit is likely due to the fact that the extragalactic FIR prescription for SFRs is appropriate for star formation timescales of 10-100 Myrs  and a well sampled IMF at high stellar masses while the SFRs for the local cloud sample are derived for a 2 Myr timescale and for a young stellar population that does not as completely sample the high mass end of the IMF. Nonetheless, these considerations suggest that at least some  upward adjustment of the GS04 SFRs may be necessary for comparison with local clouds. 

 Another consequence of the upward adjustment of the star formation rates is that of a corresponding  decrease in the estimated total molecular gas depletion times for the GS04 galaxies. This decrease would amount to a factor of 2.7 for the adjustment factor we adopted and have potentially important consequences for our understanding of galaxy evolution. These decreased gas depletion times for the GS04 galaxies are consistent those that describe the local galactic clouds (e.g., Figure 1).  However, we hesitate in drawing too firm a conclusion regarding this particular issue since it does depend somewhat on our choice of adjustment options (i.,e., a, b, or c). For example, if we selected option (b) above, only the depletion time for the dense gas component of the galaxies would be lowered. It is also interesting to note in this context that the depletion times for the dense star-forming gas are typically an order of magnitude lower than those estimated for the total molecular gas component in both galaxies and local clouds, and this remains true independent of any adjustments to the galaxy data. 
 
 
As discussed earlier, instead of adjusting the star formation rates, we could have adjusted the GS04 galaxy masses (downward) by the same constant offset in log(M). By not correcting the mass estimates we are assuming that the molecular-line derived  masses and the extinction derived masses accurately reflect the same cloud material, that is, $M_{DG}$ $=$ $M_{HCN}$ and $M_{TG}$ $=$ $M_{CO}$. To assess this possibility for the case of the total cloud masses, $M_{TG}$, we compared the extinction measurements with CO observations of a subset of the local cloud sample. We obtained CO data for five of the clouds from the archive of the CfA 1.2 m Millimeter-wave Telescope (Dame et al. 2001). The $^{12}$CO observations were averaged over the individual clouds and  the integrated CO intensities were measured for each cloud. Applying the standard CO-to-H$_2$ conversion factor of 2 x 10$^{20}$ cm$^{-2}$ (K km s$^{-1}$)$^{-1}$ (Dame et al. 2001) to convert the integrated intensities to H$_2$ column densities, we determined the mass of each cloud. We found these CO derived masses to all agree with the corresponding extinction (A$_K$ $\geq$ 0.1 mag) derived masses to better than 12\%, indicating that the extinction (A$_K$ $>$ 0.1 mag) and CO derived total masses both trace the same cloud material for local clouds. This suggests that total masses derived from CO can be a good proxy for extinction derived total masses and thus that the masses derived from CO observations of galaxies can be compared directly with those of the local cloud sample, provided that  the galaxy measurements trace the summed CO emission from a population of GMCs. If there is any diffuse CO emission from inert, non star-forming, molecular gas contributing to the galaxy-averaged CO measurements, then the CO masses derived for galaxies overestimate the masses in star forming GMCs.  In such a case the CO derived masses for the galaxies would have to be adjusted downwards to compare to the local observations. 

 A similar comparison of extinction and HCN derived masses is not possible for the local clouds since the corresponding HCN observations of these clouds do not exist. This is unfortunate because the HCN masses derived by GS04 are likely upper limits to the true masses (Gao and Solomon 2004b). For example, if the clouds are bound but not virialized then the derived masses could be somewhat underestimated. Thus, although it appears that the extragalactic CO derived masses can be directly placed on the SFR-Molecular Cloud Mass diagram without any systematic adjustment, the situation is somewhat  less certain for the HCN masses derived by GS04. However,   we note that the average ratio of dense gas (i.e., A$_V$ $\geq$ 0.8 mag) to total cloud mass (i.e., A$_V$ $\geq$ 0.1 mag.) calculated from the extinction data is $<f_{DG}>$ $=$ 0.10 $\pm$ 0.06 for the  sample of local clouds. For the GS04 sample of galaxies we find $<f_{DG}>$ = 0.16 $\pm$ 0.14 comparing the HCN and CO derived masses. The relatively close correspondence of $f_{DG}$ for these two samples is consistent with the idea  that the high extinction and HCN observations trace roughly similar fractions of the total cloud masses and thus similar dense material in clouds and galaxies, (i.e., $M_{DG} = M_{HCN}$).  This suggests that  the extragalactic HCN and CO  observations of Gao and Solomon likely trace similar material as observed in the extinction observations of Galactic clouds by Lada et al.(2010) and thus both sets of masses can be directly placed on the SFR-Molecular Mass diagram without systematic adjustment.


We note here that instead of plotting galaxies on the SFR--Molecular Mass diagram many authors traditionally prefer to plot them on the $\Sigma_{\rm{SFR}}$ -- $\Sigma_{gas}$ diagram, arguing that these two latter quantities are not affected by the (correlated) errors induced by inaccuracies in the galaxy distance measurements.  However, we prefer to plot the total formation rate, SFR,  as function of the gas mass, $M_G$, to better compare the local sample with the extragalactic one. In doing so, we acknowledge the fact that the distance-squared factor entering the evaluation of the total mass and total star-formation rate could induce a potentially strong correlation between these two variables.  This correlation, in turn, might hide the real power-law index of the underlying relation, making it appear closer to unity than it is in reality (this is a consequence of the fact that the distance enters with the same exponent, two, in both quantities).  On the other hand, simple statistical arguments and numerical checks show that the measured slope of the relation is significantly biased only in the limit where the relative error on the square of the distance is of the same order of magnitude, or larger, than the range spanned by the data.  In our case, the extragalactic data set spans approximately 4 orders of magnitude, and distances errors are on the order of $30\%$ or less, and therefore we are affected by a negligible bias in the measurement of the slope of the underlying relation using the SFR-Molecular Mass diagram.


\section{Discussion and Conclusions}

The SFR-Molecular Mass diagram of Figure 2 provides a physical context for understanding the star formation scaling laws over spatial scales ranging from those of local molecular clouds to those of entire galaxies. The  close correlation of the star formation rate with the mass of dense gas over these immense scales has been established in previous studies (Wu et al. 2005, Paper I).  Here we find that a close relation also appears to hold between the SFR and the total molecular mass over a similarly large range, 8-9 orders of magnitude in both quantities.  Both the local clouds and galaxies appear to scatter around the linear relation given by Equation 1 for $f_{DG}$ $=$ 0.1 and $M_G = M_{TG}$.  From extrapolation of the results for local clouds we suggest that this particular line corresponds to the case where 10\% of the measured gas mass is in the form of dense, star forming material for the galaxies as well as for the local clouds. The smaller scatter of the galaxies around this relation compared to that of the local clouds is likely the result of the fact that the galaxy measurements are averages over entire systems. 

These results indicate that, similar to the situation for dense gas, the star formation scaling law for total (H$_2$ + He) gas mass is likely linear across all scales for molecular clouds with similar dense gas fractions.  This notion is reinforced by the recent observations of Daddi et al. (2010) who studied infrared-selected BzK galaxies at $z\sim1.5$ and found evidence for unusually high gas fractions and extended molecular reservoirs in these distant systems. Using the star formation rates and CO gas masses provided by Daddi et al. (2010), we plot these six galaxies (open triangles) on Figure 2 and find that the BzK galaxies occupy an area in the SFR-Molecular Mass plot that is close to the linear relation described by Equation 1, consistent with the locations of Gao and Solomon galaxies and the extrapolation of the local Galactic cloud sample. 

These results lead us to the conclusion that there is a basic and universal physical process of star formation that presently operates in our Milky Way galaxy and is also responsible for the bulk of star forming activity occurring in external galaxies both in the present epoch (z $\approx$ 0; GS04) and perhaps at much earlier (z $\approx 1-2$; Daddi et al. 2010) cosmic times. It is a process in which the rate of star formation is simply and directly controlled by the amount of dense molecular gas that can be assembled within a star forming complex. In most situations massive molecular clouds appear to be able to convert only about 10\% or less of their total mass into a sufficiently dense (n(H$_2$) $\ge$ 10$^4$ \cc) form to actively produce stars. This may be considered as the normal process of star formation in GMCs. 

Closer inspection of Figure 2 suggests that for starburst galaxies, particularly the ULIRGS, this standard scenario may be modified. As the SFRs for starbursts (i.e., LIRGs and ULIRGs in Figure 2) increase with gas mass, the open symbols (CO derived gas masses) appear to approach and then merge with the filled symbols (HCN derived gas masses), almost  overlapping at the highest SFRs. As originally hypothesized by Gao and Solomon (2004), we interpret this to indicate that these galaxies are characterized by an increasingly  high dense gas fraction and consequently, the CO observations begin to trace nearly the same material as the HCN observations. Nonetheless, the star formation rate is still dictated by the amount of dense gas within the galaxies. This interpretation is also favored by Heiderman et al. (2010) who suggested that the maximal starburst activity occurs when $f_{DG} = \rm 1$ which they posit to happen when the mass surface density exceeds values $\sim$ 10$^4$ \msun\ pc$^{-2}$. ULIRGS (e.g.,  Arp 200)  are believed to be experiencing major mergers and we suggest that this extreme process likely produces conditions (e.g., high pressures) that could increase the dense gas fractions of the molecular clouds within these systems (e.g., Blitz \& Rosolowsky 2006).  In contrast the BzK galaxies studied by Daddi et al. (2010) have similarly high SFRs but lower dense gas fractions. Their high star formation rates appear to result from high global molecular gas mass fractions (i.e., M$_{H_2}$/M$_*$), as might be expected for very young galaxies.

We note that a linear relation in the SFR-Mass plane should transform to a linear relation in the $\Sigma_{SFR}$-$\Sigma_{g}$ plane (provided the surface densities for the galaxies are global averages) and we can express our star formation scaling law in this latter plane as:

\begin{equation}
\Sigma_{SFR} \propto f_{DG}\Sigma_{g}
\end{equation}

\noindent
where $\Sigma_{g}$ refers to the H$_2$ gas mass. 
Moreover, the Spitzer study of Galactic clouds by Heiderman et al. (2010) suggested a linear star formation law in the $\Sigma_{SFR}$-$\Sigma_{g}$ plane that holds for gas above a threshold surface density of $\sim$ 130 \msun\ pc$^{-2}$ (i.e., A$_K$ $>$ 0.9 mag) and extrapolates smoothly to the GS04 galaxies. 

Our result  is apparently not consistent with the standard Schmidt-Kennicutt, super-linear, scaling law (Kennicutt 1998a). Both are based on valid empirical relations. However, here we argue that the underlying scaling law for star formation is linear over all scales for all clouds and galaxies, provided they are characterized by the same dense gas fraction. Kennicutt (1998a) uses total (HI $+$ H$_2$) gas mass surface densities with CO derived  molecular masses and combines results for both normal star-forming disk galaxies and starburst galaxies to derive his star formation scaling law. Note that for these latter galaxies the total gas surface densities are dominated by the molecular component. The starbursts dominate his relation for $\Sigma_{gas}$ $>$ 100 \msun pc$^{-2}$. It is possible that the fit of a single relation to the combined sample with CO determined masses is inappropriate and skewed by the starbursts because $f_{DG}$ for starbursts is higher than that for normal star forming spirals. Indeed, Gao and Solomon (2004) showed that using the masses calculated from the CO observations  produced a super-linear (n $\approx$ 1.7) scaling law (in the SFR vs M$_G$ plane) for a sample that included their galaxies and an additional number of luminous starbursts drawn form the literature. Using gas masses derived solely from HCN observations, however, produces a linear star formation law connecting both normal star forming galaxies and starbursts. The standard Schmidt-Kennicutt relation may also be skewed at low mass surface densities. For galaxies in this portion of the diagram, the HI surface density is a large fraction of the total gas surface density and thus the measured total gas surface density, $\Sigma_{HI + H_2}$, contains a large component of inert, non-star forming, (HI) gas; this dilutes and lowers the SFR corresponding to a fixed mass surface density, resulting in a steepening of the slope of the $\Sigma_{SFR}$ vs $\Sigma_{gas}$ relation.  

These two effects, the increasing dense gas fraction for the starbursts and the dilution of the dense gas fraction by HI gas at low gas surface densities, which act together to steepen the slope of the Schmidt-Kennicutt relation, can also explain the finding of Heiderman et al. (2010) and Evans et al. (2009) that the extrapolation of the extragalactic scaling relations to local scales (i.e., low mass surface densities) lies below the data for Galactic clouds. 
    
  It can also be shown that our scaling law (Equation 1) is consistent with a volumetric scaling law, $\dot{\rho_*}  \propto \rho_G^n$ if and only if $n = 1$ and $\rho_G \geq \rho_{thres}$, where  $\rho_{thres}/\mu$ corresponds to the threshold volumetric number density for star formation  for a mean particle mass given by $\mu$ (i.e., n(H$_2$) $\geq$ 10$^4$ cm$^{-3}$).

As discussed earlier, taking the empirical, linear star-forming scaling relations at face value leads to a simple interpretation of the observations in Figures 1 \& 2. Namely, that the total rate of star formation, $\dot{M}_*$, is directly proportional to the mass of dense molecular gas above a threshold density, $M_{DG} = \int_{\rho_{thres}}^\infty M(\rho) d\rho$.  Moreover, once the gas has reached this threshold density, the SFR does not depend on the exact value of the density but only on the total mass of gas whose density has exceeded the threshold. This interpretation of the observations differs from those that explain the observed non-linear index of the Schmidt-Kennicutt law as resulting from star formation timescales dictated by the free-fall time, e.g.,  SFR $\sim$ M/$\tau_{ff}$ $\sim$ $\rho/ \rho^{-0.5}$ $\sim$ $\rho^{1.5}$ since $\tau_{ff} \sim \rho^{-0.5}$ (e.g., Elmegreen 1994; Krumholz \& Thompson 2007; Narayanan et al. 2008). A recent variant of such a model has been studied by Krumholz et al. (2011). They propose that the underlying physical law governing the relation between star formation rates and cloud properties is given by  $\dot{\rho_*}  \propto \rho_G/ \tau_{ff}$. They find that the standard Schmidt-Kennicutt law can be linearized if the data are plotted in the $\Sigma_{SFR}-\Sigma_G/\tau_{ff}$ plane as long as the free-fall times are measured using the typically higher densities of the star forming gas rather than those derived from the mean densities averaged over kpc scales. Their interpretation differs from the one in this paper in that Krumholz et al. (2011) posit that the positions of galaxies in the standard $\Sigma_{SFR}$-$\Sigma_{G}$ plane are a consequence of both their gas surface densities and their local free-fall times, while here we posit that the locations of these galaxies instead depend on their gas surface densities and their dense gas fractions. Although both interpretations are consistent with the observations, they appear not to be consistent with each other. However, we point out that Figure 1 empirically demonstrates that the locations of Galactic clouds in the SFR-Mass diagram are in fact a result of their dense gas fractions. Therefore it seems reasonable to infer that the locations of galaxies in the diagram are due to the same cause.

Finally, we reiterate  our point that the linear scaling law of Equation 1 implies that the process of star formation across entire galaxies as well as individual local clouds is governed by a very similar and simple physical principle: the rate at which molecular gas is turned into stars depends on the mass of dense gas within a molecular cloud or cloud population. The underlying star formation scaling law is linear over all scales for all clouds and galaxies characterized by  the same dense gas fraction. The star formation rate appears therefore to be controlled by local processes and not by global, galactic scale mechanisms, except to the extent that such mechanisms can alter the dense gas fractions in the molecular gas. If this interpretation is correct, then the key problem that needs to be addressed in future studies is that of the origin of the dense gas component of molecular clouds.


\acknowledgments

We thank Leo Blitz, Tom Dame, Daniela Calzetti, Bruce Elmegreen, Debbie Elmegreen,  Neal Evans, Reinhard Genzel, and Mark Krumholz for informative discussions and comments and Tom Dame for providing us with CO data.

\clearpage



\clearpage

\begin{figure}
\includegraphics[width=\linewidth, bb=50 50 280 275]{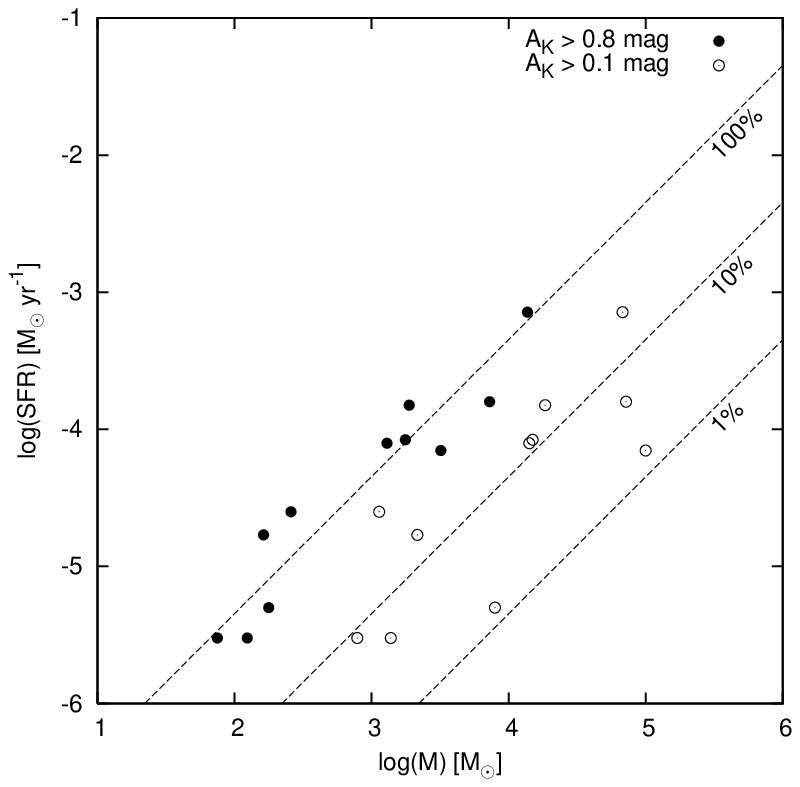}
\caption{The SFR-Molecular Mass diagram for local molecular clouds. The solid symbols indicate cloud masses above an extinction threshold of  A$_K$ $=$ 0.8 magnitudes (dense gas masses) while open circles correspond to cloud masses above A$_K$ $=$ 0.1 magnitudes (total cloud masses). The parallel dashed lines are linear relations that indicate constant fractions of dense (i.e., A$_K$ $\geq$ 0.8 magnitudes; n(H$_2$) $\geq$ 10$^4$ \cc) gas. The top line is the best linear fit to the solid symbols and represents  the case where all the gas measured is dense star forming material. (see text). \label{fig1}}
\end{figure}

\begin{figure}
\vskip -0.15in
\includegraphics[width=\linewidth, bb=50 50 280 275]{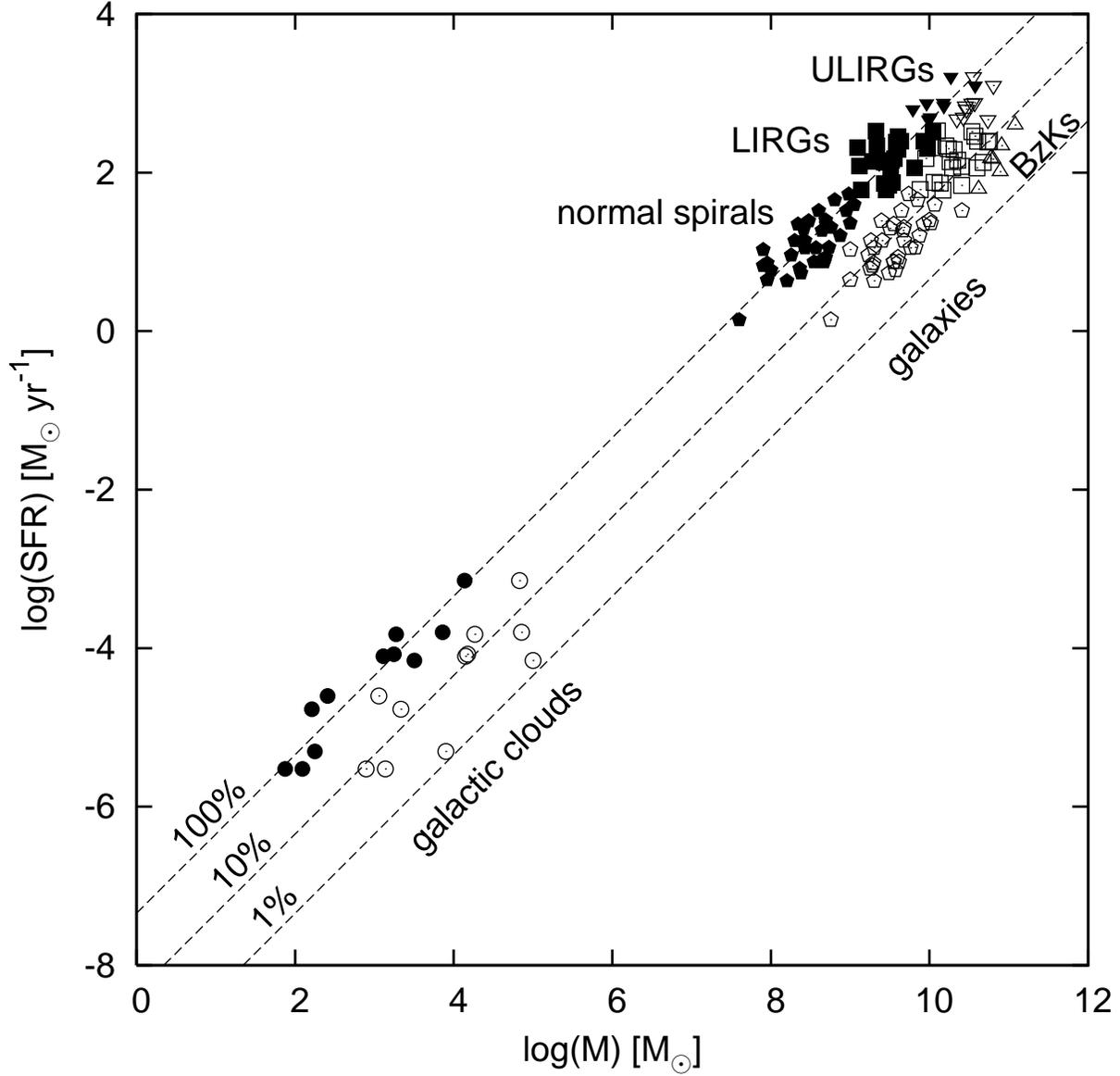}
\vskip -0.3in
\caption{The SFR-Molecular Mass diagram for local molecular clouds and galaxies from the Gao and Solomon (2004) sample. The solid symbols correspond to measurements of dense cloud masses either from extinction observations of the galactic clouds or HCN observations of the galaxies. The open symbols correspond to measurements of total cloud masses of the same clouds and galaxies, either from extinction measurements for the galactic clouds or CO observations for the galaxies.  For the galaxies, pentagons represent the locations of normal spirals, while the positions of starburst galaxies are represented by squares (LIRGS) and inverted triangles (ULIRGS). Triangles represent high-z BzK galaxies. The star formation rates for the Gao and Solomon galaxies have been adjusted upward by a factor of 2.7 to match those of galactic clouds when extrapolated to local cloud masses.  (see text). \label{fig1}}
\end{figure}





\end{document}